\documentclass[a4paper,fleqn]{cas-dc}

\usepackage{paralist}
\usepackage{enumitem}
\usepackage{caption}
\usepackage{subcaption}
\usepackage{soul}
\usepackage{mathtools}
\usepackage{amsmath}
\usepackage[authoryear,longnamesfirst]{natbib}

\def\tsc#1{\csdef{#1}{\textsc{\lowercase{#1}}\xspace}}
\tsc{WGM}
\tsc{QE}
\tsc{EP}
\tsc{PMS}
\tsc{BEC}
\tsc{DE}

\begin{document}
\let\WriteBookmarks\relax
\def\floatpagepagefraction{1}
\def\textpagefraction{.001}

\shorttitle{Low-Frequency Tweeters Have More to Say!}

\shortauthors{Anonymous et~al.}

\title [mode = title]{``Low Frequency Tweeters Have More to Say!'' A New Approach to Identify Importance of Tweets}   





\author[1]{Gautam Khanna}[]
\cormark[1]

\ead{gautamkhanna@hotmail.com}



\affiliation[1]{organization={University of Manchester},
    addressline={Kilburn Building}, 
    city={Manchester},
    postcode={M13 9PL}, 
    country={United Kingdom}}

\author[2]{Yeliz Yesilada}[                     orcid=0000-0001-7511-2910]

\ead{yyeliz@metu.edu.tr}

\affiliation[2]{organization={Middle East Technical University Northern Cyprus Campus},
    city={Kalkanlı},
    postcode={99738}, 
    state={Guzelyurt},
    country={Turkey}}

    \author[2]{Sukru Eraslan}[                     orcid= 0000-0002-9277-8375]

    \ead{seraslan@metu.edu.tr}

     \author[1]{Simon Harper}[                     orcid=0000-0001-9301-5049]

     \ead{simon.harper@manchester.ac.uk}










\begin{abstract}
Twitter is one of the most popular social media platforms. With a large number of tweets, the activity feed of users becomes noisy, challenging to read, and most importantly tweets often get lost. We present a new approach to personalise the ranking of the tweets toward solving the problem of information overload which is achieved by analysing the relationship between the importance of tweets to the frequency at which the author tweets. The hypothesis tested is that ``low-frequency tweeters have more to say'', i.e. if a user who tweets infrequently actually goes to the effort of tweeting, then it is more likely to be of more importance or contain more ``meaning'' than a tweet by a user who tweets continuously. We propose six new measures to evaluate the \textit{importance} of tweets based on the ability of the tweet to drive interaction among its readers, which is measured through metrics such as retweets, favourites, and comments, and the extent of the author's network interacting with the tweet. Our study shows that users who tweeted less than ten tweets per week were more likely to be perceived as important by their followers and have the most important messages. This identified tweet-frequency band could be used to reorder the activity feed of users and such reordering would ensure the messages of low-frequency tweeters do not get lost in the stream of tweets. This could also serve as a scoring index for Twitter users to identify users frequently tweeting important messages.
\end{abstract}






\makeatletter\def\Hy@Warning#1{}\makeatother

\maketitle

\section{Introduction}\label{sec:intro}
Twitter is one of the most popular social media platforms with an active user base of more than 330 million worldwide~\citep{StatistaTwitterStatista}. 
%
This extensive use of Twitter has led to an overabundance of tweets, causing users' activity feed to become noisy and difficult to read. 
Often in the plethora of tweets, important messages get lost. The primary motivation of our study is to find a solution to this problem of information overload on Twitter by personalising the ranking of the tweets to the users and helping them reduce the effort to find the required information.
The hypothesis tested is ``low-frequency tweeters have more to say''; i.e. if a user who tweets infrequently actually goes to the effort of tweeting, then it is more likely to be of more importance or contain more \textit{meaning} than a tweet by a user who tweets continuously. The study was carried out with the objective of finding the solution to two research questions: (1) ``How can you distinguish important tweets?'' (2) ``Is there any correlation between frequency of tweeting activity
and importance of tweet?''. The emphasis was to analyse the relationship between the importance of a tweet and the frequency at which the author tweets and determine whether there is any correlation between the tweet frequency of a user to the importance of their tweets and if the tweet frequency of the author can be effectively used as a metric for the potential importance of tweet. The research thus aims to establish the author as a potential indicator of the importance of a tweet rather than the contents tweet itself.

Although there has been significant research on the ranking of Twitter users and tweets (Section \ref{sec:related}), they have primarily been aimed at identifying `Influential' user and their network impact to improve viral marketing campaigns. The `importance' of tweet and user in this context thus often is defined by the ability of the tweet/user to reach the maximum audience. However, our research here focuses on the issue of information overload and is aimed at providing relevant information to the reader without consideration of the extent of the reach of the user or tweet.

The study was carried out by selecting a set of (pseudo) random UK non-verified Twitter users based on selection criteria from the active user base. Our study has three parts: the Formative experiment and the Summative experiment, and validation with celebrity accounts. 
In the formative experiment, we establish the importance of a tweet and its author (see Section~\ref{sec:formative}). In this research, the importance of a tweet is measured by the ability of any tweet to engage and interact with the audience. It further considers the extent of the audience that has interacted with the tweet. Users can interact with tweets through retweets, favourites, comments, and quoted tweets. Two tweet--level and four users--level measures of importance were proposed in the research to evaluate the Tweets and their authors and determine their level of importance. The proposed measures of the importance of a tweet and its author consider these metrics while factoring in the extent of interaction, i.e., the percentage of the original audience (the author's followers) that interacted with the tweet. 

In the summative experiment, we validate our proposed metrics (Section~\ref{sec:summative}). The results obtained on the test set of tweets for the sample were evaluated against a validation set of tweets for the same sample of users. Further, the results were evaluated against selected verified or celebrity accounts.
Through the frequency analysis of the tweet importance metrics proposed in the formative experiment, it was established that low-frequency tweeters were more likely to be of higher importance than those tweeting with a high frequency. It was determined that users tweeting less than ten tweets per week were more likely to be perceived as important by their followers (Section~\ref{sec:discussion} and~\ref{sec:conclusion}). 

The main \textbf{contribution} of this study is showing that users who tweeted less than ten tweets per week were more likely to be perceived as important by their followers and have the most important messages. This identified tweet-frequency band could be used for different purposes. It could be used to reorder the activity feed of users and such reordering would ensure the messages of low-frequency tweeters do not get lost in the stream of tweets. This could also serve as a scoring index for Twitter users to identify users frequently tweeting important messages.


\section{Related Work}\label{sec:related}
Two lines of research are related to our work: 1) identifying the `influential' users or ranking/importance of the users, and 2) identifying the ranking/importance of the tweets. 
Although there is significant research on both, there is no previous work investigating the importance of a tweet based on the frequency of the tweeting activity of the users. There are also studies in understanding the factors affecting the frequency of use~\citep{LAOR2022101922,doi:10.1080/1461670X.2014.939849} but in our work we do not aim to predict the frequency of the use. 

Measures are introduced in analysing the influence of Twitter users~\citep{Zhu_2022}; for instance, \cite{Cha2010MeasuringFallacy} propose in-degree, retweet, and mentions as measures and suggests that the in-degree of any user (i.e. the number of followers) represents the popularity of the users and is not related to their ability to influence or engage their audience. 
Various approaches are also explored in the literature to identify influential users~\citep{10.1145/3155897}. 
Community Scale-Sensitive Maxdegree (CSSM) algorithm is proposed to discover users with the maximum potential influence to promote products~\citep{Hao2012DiscoveringMechanism}. The algorithm aims to discover the most powerful nodes to maximise influence. 
The proposed approach improves upon the natural greedy algorithm~\citep{Kempe2003MaximizingNetwork}, and its derivative approaches.~\cite{Kundu2011ANetworks} propose a centrality measure for the problem of influence maximisation in social networks. 
~\cite{ChenEfficientNetworks} approach the problem of efficient influence maximisation by improving the original greedy approach and its improvements to reduce running time and by proposing a degree discount heuristics that improve influence spread. Although the CCSM algorithm outperforms the previous methods, it approaches the problem of influence using a traditional approach considering only the number of users the message is shared with while not considering the perception of the audience towards the nodal user as it does not consider the ability of the user to engage its audience.~\cite{Yamaguchi2010} also propose a ranking algorithm for Twitter users called TURank that evaluates Twitter users' authority score on link analysis. 
~\cite{BalminObjectRank:} apply this to evaluate users' authority scores. 
The algorithm focuses on the post, retweets, and follows the behaviour of users. 
As the algorithm considers users to be an authoritative user on the absolute number of `useful tweets' posted by the user, it does not consider the non-useful tweets posted by the user while ranking them. Thus, it would be biased towards users who share many messages with a fraction of their messages perceived as important by the reader, against those who share fewer messages but a high majority of which are considered as important by the reader. Similar to this approach,~\cite{10.1007/978-3-642-25853-4_11} propose a tweet-centric approach for the topic-specific ranking of Twitter users. 
This method also considers the absolute number of tweets while computing the rank of the author, thus inherently containing the same bias as the previous approach. \cite{Jianqiang2017} propose an algorithm called UIRank, which identifies influential users through interaction information flow and interaction relationships among users in the micro-blog. The influence of a tweet is described by its ability to cause a change in opinion, behaviour, or emotion in the reader. The algorithm considers two factors to determine the influence of a user: the influence of tweets and commenting behaviour. 
The study provides valuable insight into measuring the ability of a tweet to influence its readers, although it does not consider the ``Like'' or ``Favourite'' activity while measuring the influence of tweets.

Besides the ranking of the users, there is also work on understanding tweet semantics and impact~\citep{10.1007/s13278-022-00872-1} and ranking the tweets.~\cite{Yao2016LeveragingGeneration} propose an optimised timeline-generating framework leveraging tweet ranking algorithms. 
The framework constructs a semantic graph based on three factors: novelty, relevance, and coverage. 
The tweets are then ranked based on the Tweet ranking model proposed which is based on the textRank~\citep{MihalceaTextRank:Texts} algorithm, to rank the tweets identified by the semantic graph. 
On the other hand, tweet ranking algorithms lean towards ranking the tweets by creating predictive models based on tweet-centric metrics such as retweets, contents, and author-based features rather than user graphs. Often the focus has been on retweeting behaviour of users presented a learning approach based on 15 tweet content-based features (e.g., the presence of hashtags, exclamations, URLs, etc.). Although their experiment shows the results obtained are not random, the model does not consider the characteristics of the author while predicting the retweeting probability. 
\cite{Uysal2011} propose a model based on Coordinate Ascent learning to rank the incoming tweets to help the users interact with the tweets they are more likely to retweet. A similar approach is also proposed by \citep{Yang2010UnderstandingNetworks}, which considers 22 features and built a retweet prediction model based on the factor graph model. Additionally,~\cite{HongOvidiuDanBrianDavisonPredictingTwitter} consider the content, graph topology, temporal, and metadata features of a tweet and built a Logistic regression model for prediction. Although these approaches aim to predict the probability of retweets from a global perspective, the approach proposed by~\citep{Uysal2011} considered an author-based, tweet-based, content-based, and user-based set of features to train the model for ranking tweets and users who were more likely to retweet. 
Furthermore, a variant of the Hyperlink-Induced Topic Search (HITS)~\citep{Kleinberg1997Authoritative} algorithm is proposed~\citep{Yang2012FindingAnalysis} for producing a static ranking of tweets based on Retweet graph analysis. 
~\cite{Kuang2016} also propose a ranking model which considers the tweet's popularity, the intimacy between the user and the author, and the field of interest for the user. The model uses metrics such as the number of retweets, comments, and likes to determine the popularity of tweets. 
Although this model also uses the tweet's popularity as a metric to rank the tweets, there is an inherent latency in the algorithm to allow the tweet to achieve its maximum retweets and thus cannot be applied to real-time tweets. 

\section{Formative Experiments}\label{sec:formative}
To test our hypothesis, we first conducted formative experiments as detailed here and then summative experiments as detailed in the following section. All of our studies received the relevant approval by the \textit{anonymised} Ethics Review Panel (Ref:2018-3969-5604).

\subsection{Participants}\label{Form:Part}
A sample of the active population was selected comprising of random users who had a public account and should:
\begin{inparaenum}[1)]
    \item Be a UK user to avoid cultural bias;
    \item Have at least one tweet in the last 30 days to eliminate dormant users;
    \item Be a non-verified account such that the users are only evaluated on the basis of their activity on Twitter, i.e., should not be a celebrity account to ensure that there are minimum external biases such as the release of an album or a political campaign; 
    \item Have more than ten tweets to provide enough sample and be statistically significant;
    \item Not be a spam account. Based on the literature \citep{VermaScholar2014,Benevenuto2010,Martinez-Romo2013,AmleshwaramCATS:Spammers,ThomasOpenAbuse,Gurajala2015,Lee2010UncoveringLearning,Gupta2012CredibilityEvents}, accounts should have the following: (i) Spam accounts are frequently created and deleted therefore the account should be older than three months (90 days); (ii) Spam accounts do not have a large number of followers and therefore account should have at least ten followers; (iii) The ratio of a number of followers to the number of following is very close to zero and therefore the number of following should not be higher than 20 times the number of followers; (iv) Spam account often have missing account characteristics such as account image, description, language and personal details, and therefore the account should have non-default values for all primary account characteristics.
\end{inparaenum}

The sample size required was calculated considering a Confidence Level of 99\% with a Confidence Interval of 1.80 for a total population of 17 million active UK Twitter users~\citep{StatistaTwitterStatistab}. The result obtained was 5135 users which were approximated to 5200 users.
A pseudo-random mechanism was adopted for user selection to overcome these issues. The setup was divided into two phases: initial and final sample. 
The initial sample of eligible users was selected over a period of one week. The tweets that were being published in the UK area were streamed live using the official Twitter Streaming API. The stream was captured for 10 minutes. The unique users who published the tweets were then filtered using the selection criteria and spam checks. The resultant set of users was added to the initial sample. This process was repeated every 60 minutes for one week. 
Further, selecting users in a 10-minute window with a 60-minute delay facilitated avoiding bias in user selection by the preferred time of activity of a user; and the total number of users selected in the initial sample after applying the selection criteria was 86557.
The final set of users was selected by generating 5200 random numbers between 1 and 86557, which served as ids for the users from the initial data set.
\subsection{Materials and Tools}\label{Form:Mat}
For the selected 5200 users, the historical tweets dataset was retrieved. As per the limitations set by the API, a maximum of the last 3200 tweets can be retrieved for any given user. This restriction includes both original tweets and retweets, although does not contain any restriction on the time since the tweet was posted. Thus, for users frequently tweeting and retweeting the dataset available corresponds to a shorter duration of time compared to infrequent tweeters. Tweet attributes downloaded for the study included \textit{user\_id, tweet\_id, created\_at, text, reteweet\_count, favourite\_count, hashtag, user\_mentions, is\_quote and is\_retweet}. 
Tweets published within the last 72 hours of retrieving were removed from the dataset as per the findings by~\citep{Kong2011AMicro-Blog}, to ensure a lifetime of all tweets considered in the study has elapsed. 
User profile attributes were also collected including \textit{user\_id, age, followers, friends, status\_count, favourite\_count}. The Twitter API does not allow to collect of historical values of User profile attributes such as some followers/following. 

    \begin{figure*}[ht]
        \centering
        \begin{subfigure}[b]{0.45\textwidth}
            \centering
            \includegraphics[width=\textwidth]{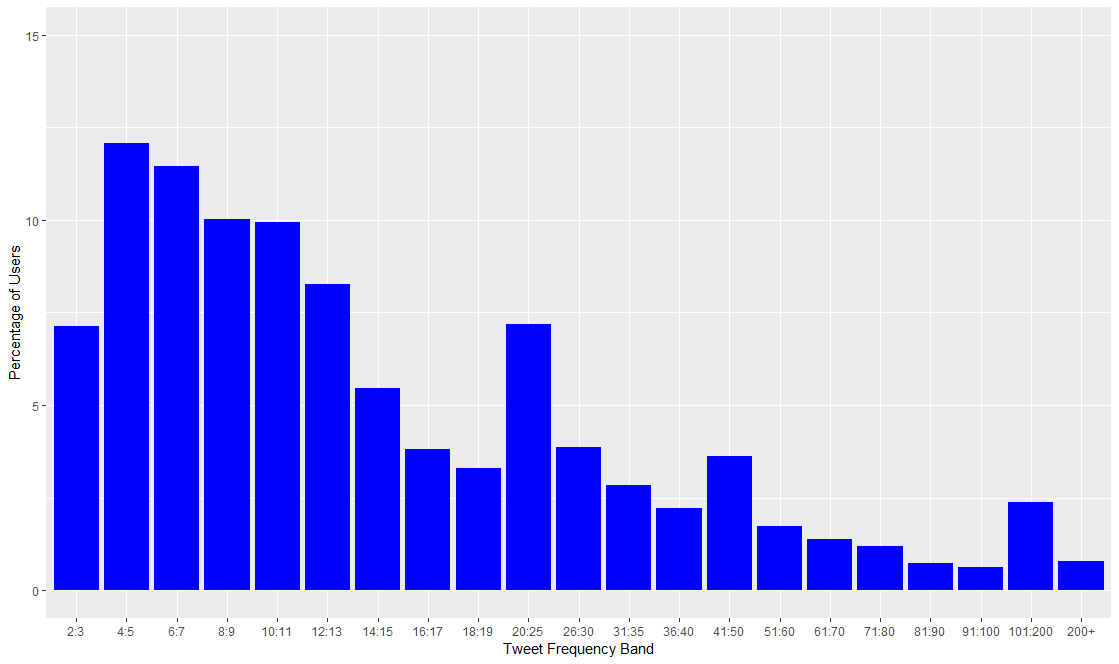}
            \caption[Frequency Distribution]%
            {{\small Frequency Distribution}}    
            \label{Figure:10}
        \end{subfigure}
        \hfil
        \begin{subfigure}[b]{0.45\textwidth}  
            \centering 
            \includegraphics[width=\textwidth]{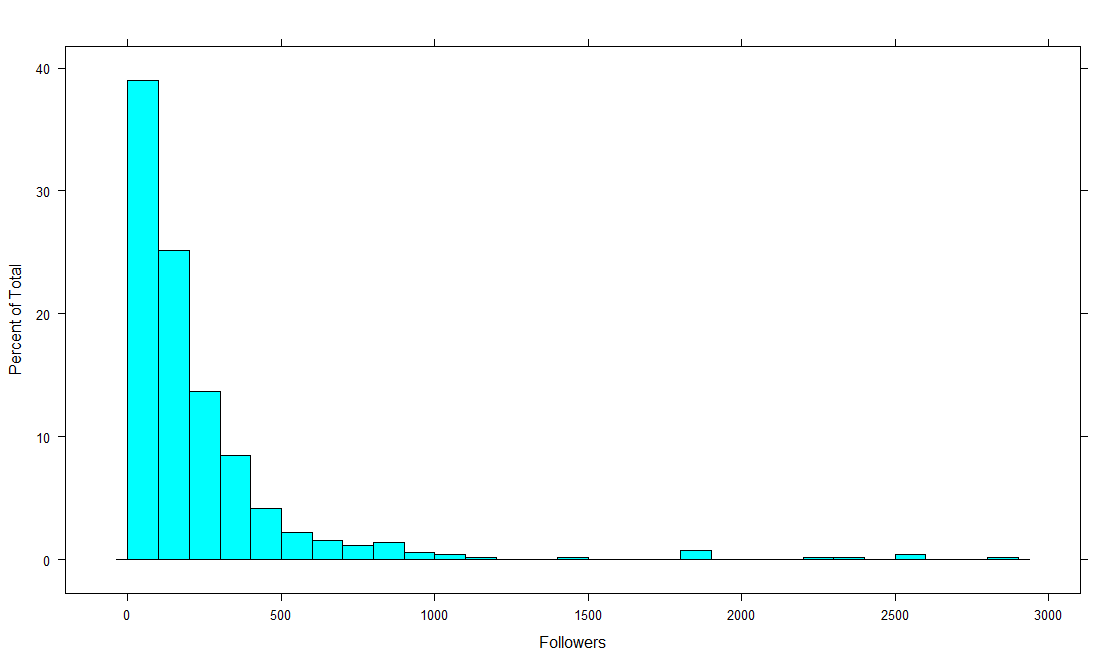}
            \caption[Follower Distribution]%
            {{\small Follower Distribution}}    
            \label{Figure:14}
        \end{subfigure}
        \caption[ Frequency \& Follower Distributions ]
        {\small Frequency \& Follower Distributions} 
        \label{Figure:freqfoll}
    \end{figure*}
\subsection{Experimental Design}\label{Form:Exp}
Two sets of metrics are used in our study: tweet level for measuring the quality of individual tweets and user level for evaluating the perceived importance of users based on their followers. 
\noindent\subsubsection*{Tweet-Level Metrics}
The quality of any given tweet was determined based on the ability of the tweet to engage its reader. User engagement can be gauged by considering the Retweet, Like, Bookmark, Comment, and Quoting. 
If a tweet can initiate any of these responses from the user, then the tweet can be considered as relevant to the reader. These tweet-level metrics help to define the user-level metrics against which a frequency analysis would be carried out to test the primary hypothesis. Table \ref{table:metrics} shows the metrics that we used.

\begin{table*}[!h]
\footnotesize
\begin{tabular}{|l|l|l|}
\hline
No & Metric                               & Meaning                                                                     \\ \hline
1  & Number of Retweets (RT)              & the absolute number of retweets.                                            \\ \hline
2  & Percentage of Retweets (prRT)        & the percentage of the initial audience retweeted.                           \\ \hline
3  & Number of Favourites (FV)            & the absolute number of times liked.                                         \\ \hline
4  & Percentage of Favourites (prFV)      & the percentage of the initial audience liked the tweet.                     \\ \hline
5  & Number of Comment/Replier (CM)       & the absolute number of   replies or comments received.                      \\ \hline
6  & Percentage of Comment/Replier (prCM) & the percentage of the initial audience commented or replied to   the tweet. \\ \hline
7  & Number of Quoted Tweets (QT)         & the absolute number of times quoted by other users.                         \\ \hline
8  & Percentage of Quoted Tweets (prQT)   & the percentage of the initial audience quoted the tweet.                    \\ \hline
9  & Number of Bookmarks (BM)             & the absolute number of times a tweet is bookmarked.                         \\ \hline
10 & Percentage of Bookmarks (prBM)       & the percentage of the initial audience bookmarked.                          \\ \hline
\end{tabular}
\caption{Metrics}
\label{table:metrics}
\end{table*}

Based on these, we introduced the following metrics:\\
\textbf{Tweet Score (TS):} The RT and FV represent the popularity of any tweet. However, they ignore the size of the initial audience of the tweet. Tweets published by users with many followers are more likely to get more retweets or likes. Similarly considering only the prRT and prFV may be biased towards users with fewer followers. Thus taking to account both factors, the TS was formulated as follows: 
\begin{equation}
\begin{aligned}
    \mathrm{TS}_{i} = (\mathrm{RT}_{i}*\mathrm{prRT}_{i})+(\mathrm{FV}_{i}*\mathrm{prFV}_{i})+(\mathrm{CM}_{i}*\mathrm{prCM}_{i})+\\(\mathrm{QT}_{i}*\mathrm{prQT}_{i})+(\mathrm{BM}_{i}*\mathrm{prBM}_{i})
    \label{eq:TS}
    \end{aligned}
\end{equation}



As the details for Replies, Bookmarks, and Quotes are not available with the Standard API, the above TS has been updated as 
\begin{equation}
    \mathrm{TS}_{i} = (\mathrm{RT}_{i}*\mathrm{prRT}_{i})+(\mathrm{FV}_{i}*\mathrm{prFV}_{i})
\end{equation}
\textbf{Tweet Score Percentile (TSPc)}: The TS provides an absolute unbounded value of the score associated with each tweet, thus making the comparison between tweets difficult. To overcome this issue, a Tweet Score Percentile value was calculated based on the TS. The TSPc value for a tweet indicates the percentage of tweets with a TS value lesser than the given tweet. Additionally, two rules were applied: (1) The TSPc value of all tweets having prRT or a prFV value greater than 100 was set to 100 (Maximum); (2) The TSPc value of all tweets having prRT or prFV value as zero was set to zero (Minimum). These rules were added to incentive tweets that engaged audiences larger than their original audience and penalise tweets that were not able to engage any user. These two cases were ignored while calculating the percentile score for tweets. 
\subsubsection*{User-level Metrics:} The perceived importance of any users was determined by evaluating the followers' behaviour in response to the tweets. If the tweets posted by any user are consistently able to engage their readers, then the author of these tweets can be considered to be perceived as an important user by its followers. The following basic user metrics were used:
\begin{inparaenum}[1)]
    \item Number of Followers -- The number of users that follow a user.
    \item Number of Friends -- The number of people the user is following.
    \item Age -- The number of days since the account was created.
    \item Status Count -- The total number of Tweets posted by the user throughout the lifetime of the account.
    \item Number Original Tweets (orT) -- The number of original tweets published by the user in the dataset.
    \item Number of Retweets (RT) -- The number of tweets retweeted by the user in the dataset.
    \item Average Original Tweets per Week (AvgOrTpW) -- The average number of original tweets posted per week by the user.
    \item Average Retweet per Week (AvgRTpW) -- The average number of Tweets retweeted per week by the user.
    \item Tweet-Frequency Band -- The users were assigned to tweet-frequency bands (see the bands in Table~\ref{table:4}) based on the AvgOrTpW. 
\end{inparaenum}

\noindent Based on these, we also computed the following metrics:\\
\textbf{Average Tweet Score (AvgTS)}: This specifies the average TS of all original tweets published by a user. This metric can be used to specify the popularity of any user as perceived by their followers. Although the metric is calculated based on TS, which is an unbounded score, the value of this score is highly volatile. A single popular tweet may increase the score of a user drastically.

\noindent\textbf{Percentage of Scored Tweets (prST)}: This specifies the percentage of original Tweets published by a user that has a positive TS, i.e., the percentage of Tweets that received at least a single like or retweet. This metric provides a measure to check how often a user can engage the audience with their tweets. 

\noindent\textbf{Average audience interaction per Week (AvgAudInpW):} Combining the attributes of the previous two metrics, AvgAudInpW is defined as the mean of the percentage of the audience that interacts with the tweets published per week (prAudInpW) as: 
\begin{equation}
\begin{aligned}
    \mathrm{prAudInpW}_{i} = \sum_{}^{}(\mathrm{RTw}_{i}+\mathrm{FVw}_{i}+\mathrm{CMw}_{i}+\mathrm{QTw}_{i}+\mathrm{BMw}_{i})\\/\mathrm{orTw}_{i}*No.of Followers
\end{aligned}
\end{equation}

Where RTw, FVw, CMw, QTw, and BMw are Retweet, Likes, Comments, Quoted Tweets, and Bookmarks received by the Tweets published by the user in a given week, orTw is the number of original tweets published by a user in a given week. Interaction with the audience is defined as the summation of the retweets, likes, comments, quoted tweets, and bookmarks received by the tweets published by the author. Similar to TS, the Comments, Quoted Tweets, and Bookmarks were unavailable due to limitations of the Standard API. The equation has been updated as follows: 
\begin{equation}
    \mathrm{prAudInpW}_{i} = \sum_{}^{}(\mathrm{RTw}_{i}+\mathrm{FVw}_{i})/\mathrm{orTw}_{i}*No.of Followers
\end{equation}
\begin{equation}
    \mathrm{AvgAudInpW}_{i} = \overline{prAudInpW}
\end{equation}

The AvgAudInpW metric serves as a measure to determine the extent of user engagement an author's tweets generate on a weekly average. As the metric does not consider the individual interaction per tweet over the entire dataset but considers the combined interaction of the tweets published within a week, it provides a holistic view of the perception of the users' importance as per their followers. By evaluating the tweets in a weekly window, the metric penalises tweeters who post tweets which do not engage the audience and promotes users that continually engage the readers. 

\noindent\textbf{Average Tweet Score Percentile (AvgTSPc):} This metric is based on the TSPc computed for each tweet. It is defined as the mean TSPc score of all tweets published by the user in the given dataset. The AvgTSPc describes the average percentile rank of the tweets published by a user, with all tweets reaching beyond the initial audience having a maximum value (100) and all tweets with no interaction with the minimum value (0). It provides a consistent measure of a user's tweets against the other users in the dataset, as it is not affected by the size of the initial audience of the tweets.

\subsection{Results}\label{sec:formativeresults}

    \begin{figure*}[ht]
        \centering
        \begin{subfigure}[b]{0.45\textwidth}
            \centering
            \includegraphics[width=\textwidth]{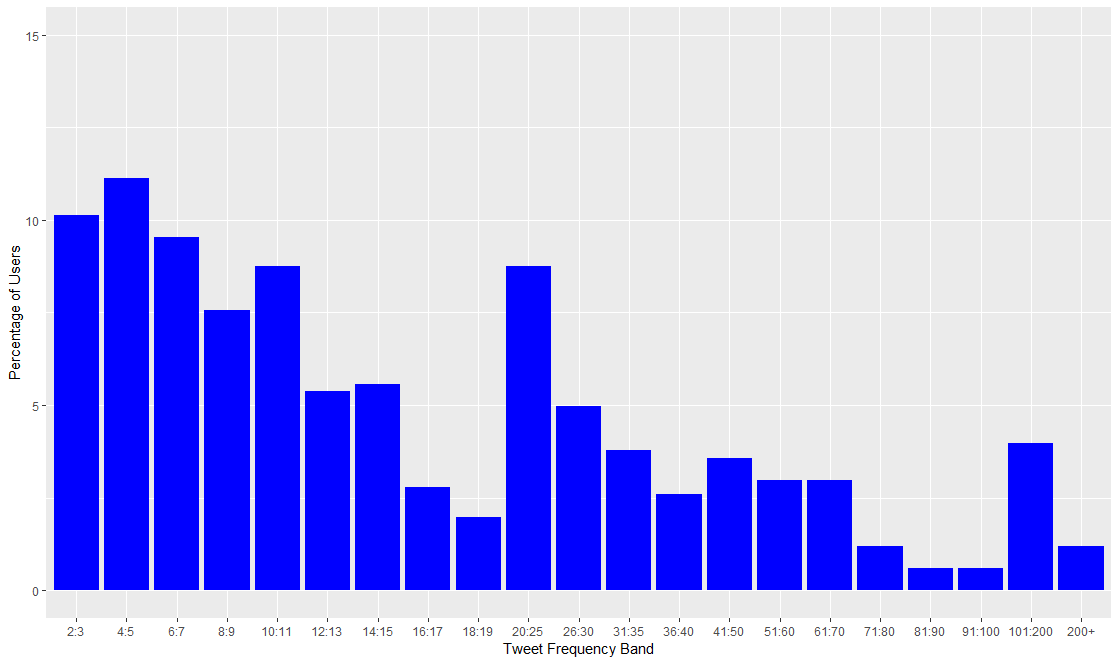}
            \caption[AvgTS -- Top Performer]%
            {{\small AvgTS -- Top Performer}}    
            \label{Figure:11}

        \end{subfigure}
        \hfil
        \begin{subfigure}[b]{0.45\textwidth}  
            \centering 
            \includegraphics[width=\textwidth]{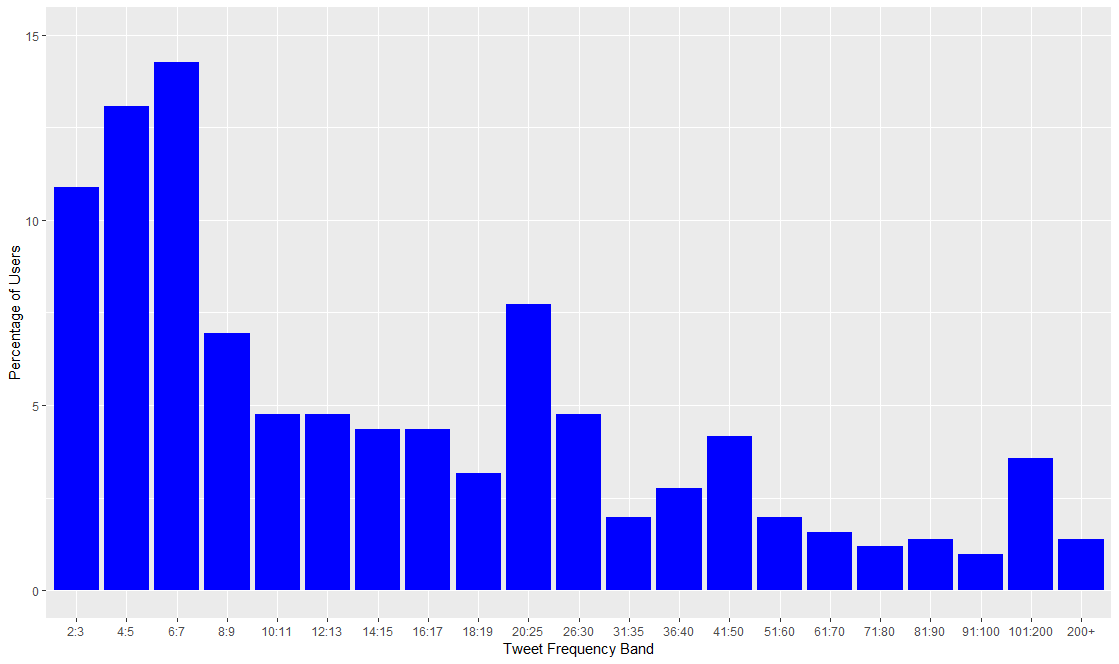}
            \caption[Percentage of Scored Tweets - Top Performer]%
            {{\small Percentage of Scored Tweets - Top Performer}}    
            \label{Figure:12}

        \end{subfigure}
        \vskip\baselineskip
        \begin{subfigure}[b]{0.45\textwidth}   
            \centering 
            \includegraphics[width=\textwidth]{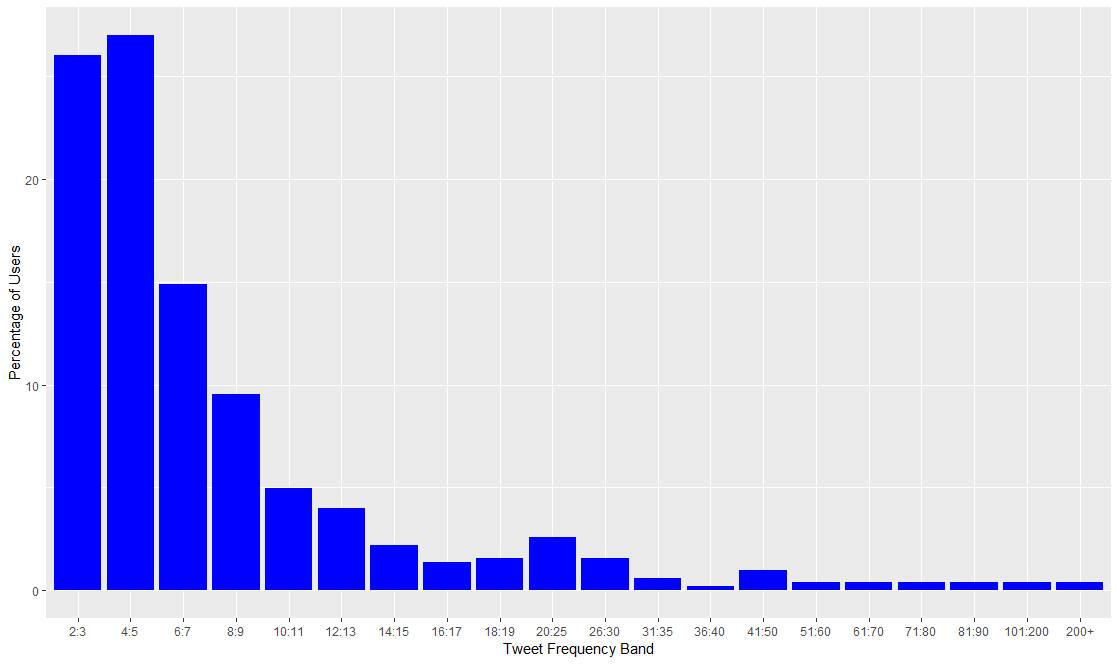}
            \caption[AvgAudInpW -- Top Performer]%
            {{\small AvgAudInpW -- Top Performer}}    
            \label{Figure:13}

        \end{subfigure}
        \hfil
        \begin{subfigure}[b]{0.45\textwidth}   
            \centering 
            \includegraphics[width=\textwidth]{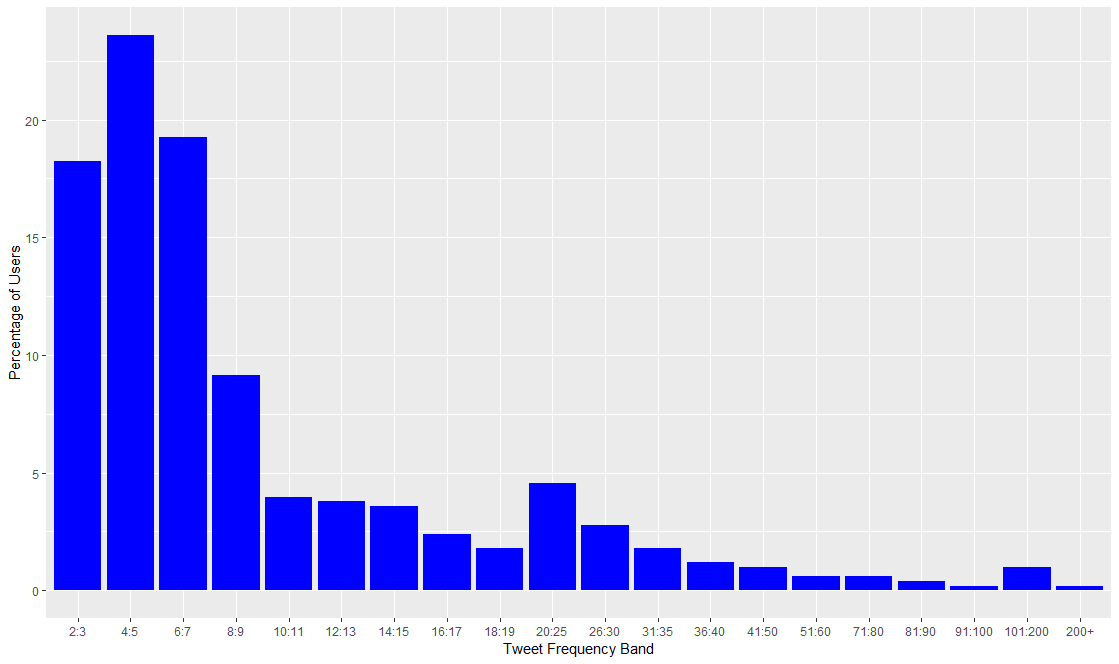}
            \caption[AvgTSPc -- Top Performer]%
            {{\small AvgTSPc -- Top Performer}}    
            \label{Figure:15}

        \end{subfigure}
        \caption[ Top Performers ]
        {\small Top Performers} 
        \label{fig:mean and std of nets2}

    \end{figure*}

A total of 5025 users were selected as the final sample. The drop in the user from 5200 random selected users was probably due to the following reasons: the user changed the profile visibility to private; the user blocked access to tweets through Twitter API; the profile was deleted between the time user was selected and Tweets were collected; the user Tweet corpus did not contain at least ten original tweets, or there were technical issues with the API.

\begin{table*}
\centering
\footnotesize
\begin{tabular}{|l|l|l|l|l|l|l|l|l|l|l|}
 \hline
 2:3 & 4:5 & 6:7 & 8:9 & 10:11 & 12:13 & 14:15 & 16:17 & 18:19 & 20:25 & 26:30 \\ \hline
 31:35 & 36:40 & 41:50 & 51:60 & 61:70 & 71:80 & 81:90 & 91:100 & 101:200 & 200+ & \\ \hline
 \end{tabular}
\caption{Tweet - Frequency Bands}
\label{table:4}
\end{table*}

The minimum age of accounts considered is 131 days, and the maximum age is 4237 days. The users are spread across different age ranges, with most accounts having an age greater than five years, although very few users have accounts older than ten years. The sample consists of user accounts ranging from accounts created at the launch of Twitter, 12 years ago, to the ones created less than six months old, covering a wide variety of users.
The users in the final sample have a wide range of followers. 
The user with the least followers had 12, whereas the user with the most had 240378. 
The average number of followers, including the outliers, is 1295 followers, whereas excluding the top 10\% of users is 616 followers.

The users were categorised into 21 frequency bands (Table~\ref{table:4}), and the distribution of users across these bands is shown in Figure~\ref{Figure:10}. The percentage of users decreases as the frequency band increases except for 20:25, 41:50, and 101:200. This is due to the change of the scale of the frequency band from the previous bands. 

The tweet-frequency band of the users were compared for the top performing users for each user-level metric to determine the relationship between Tweet-frequency and the importance of Tweet. The top 10\% scores, i.e. a user with a percentile rank of 90 or higher, were considered the top performing users for each metric.
\\The \textbf{Average Tweet Score (AvgTS)} has an unbounded numeric value, and thus, the minimum and maximum values vary significantly. The third quantile (75th percentile) and 90th percentile ranks for the score are 5.8 and 22.95, respectively. 
Figure~\ref{Figure:11} represents the tweet-frequency band distributions for 
top performer groups.

The mean number of followers for the Third Quantile group (top 25\%) is 1161, which is within range of the overall sample average. An even population distribution can be seen in the top performer group compared to the base frequency distribution. Although there is a significant increase in the population of the first frequency band, there are no other significant distribution changes. All other frequency bands have a slightly higher or lower population distribution. The mean number of followers for this group is 1516, which is significantly higher than the overall population sample. 
Since the top performer group had a smooth distribution, it can be inferred that the AvgTS is highly dependent on the number of followers of the author. Tweets with a larger initial audience are more likely to achieve a higher TS. For example, a tweet with an initial audience of 100 users, even if it engages its entire audience, would obtain a TS of 100. In contrast, a tweet with an initial audience of 100,000 engaging 10\% of its audience would obtain a TS of 10,000. Thus the top-performing group for AvgTS has a significantly more number of followers. 

The 90th percentile value for the Percentage of Scored Tweets (prST) score was 74.9\%. Figure~\ref{Figure:12} shows the frequency distribution of top performers.
Comparing the base frequency bands results, it can be seen that the first three tweet frequency bands have a substantially larger population percentage than the remaining bands. A minor increase can also be observed for the 20-25 and 101-200 bands, although the increase is considerably smaller than the first three bands.


The figure shows that lower frequency band users form a higher percentage of the top performer group for the prST metric. Although the prST considers only the percentage of tweets with a positive score, it may be possible that the top performer group comprises users primarily with few tweets in the dataset. 

The third quantile and 90th percentile scores for AvgAudInpW are 0.004 and 0.010, respectively. 
Figure~\ref{Figure:13} represents the Frequency distribution for the top performer group.


The first four frequency bands have a significantly higher population than the remaining bands. Similar to the AvgTS metrics, the AvgAudInpW metric is highly dependent on the number of followers of the author.  Where AvgTS tends towards a bias towards the users with a larger following, AvgAudInpW tends towards authors with a smaller following. Figure~\ref{Figure:14} shows the distribution of the top performer group for AvgAudInpW based on their number of followers. It can be seen that the users generally have a lower number of followers with the mean number of followers for the group as 235 and the maximum number of followers being 2856. This is due to the assumption that all tweets are read by the entire audience, as stated in the previous section. Thus authors with more considerable following tend to have a lower score as a more significant percentage of their audience may not have read their tweet.

The third and 90th percentile scores for \textbf{AvgTSPc} are 33.19 and 46.19, respectively. 
Figure~\ref{Figure:15} represents the Frequency distribution for the 90th percentile group.

The AvgTSPc provides the percentile rank based on the TS for the tweets published by the users. The 90th percentile score of 46.19 can be read as ``on an average, the tweets published by the author have received a TS better than 46.19 \% of all tweets scored in the dataset''. As seen from the figure, the first four frequency bands have a significantly higher population than the other bands.

The top performer group has a significantly larger population distribution for the first three bands, whereas the population distribution post the fourth band is extremely small (less than 5\%). This can be inferred as users tweeting up to 9 times a week on average have tweeted with a TS better than at least 46.19\% of the total tweets evaluated in the dataset. Since the AvgTSPc does not depend on the size of the initial audience, it provides a consistent measure of a user's tweets against other tweets in the dataset.

\section{Summative Experiments}\label{sec:summative}
To validate the results of the formative experiments, we also conducted summative experiments where we evaluated a dataset against the metrics identified in the formative setup. 
The \textbf{participants} were the final set of users that were evaluated for the Formative Experiment (i.e., 5025 users). 
The users were verified against the selection criteria, and spam checks were identified in Section~\ref{Form:Part}. 
Regarding the \textbf{materials}, the Tweet dataset of 5025 users for the experiment was collected following two rules: (1) All tweets published by the user within 30 days after the collection of the Tweet corpus of the Formative Experiment; and (2) the Last 3200 tweets published if the user publishes more than 3200 tweets after 30 days of the collection of the Tweet corpus of the Formative Experiment. Similar to the Formative Experiment, Tweets published within the last 72 hours of retrieving were removed from the dataset as per the findings by~\citep{Kong2011AMicro-Blog}. The same tweet and user profile attributes are collected as with the formative experiments. 
The experimental \textbf{design} with the \textbf{metrics} and the \textbf{tools} used follows the exact setup as the Formative Experiment (Section~\ref{Form:Exp}). 
\subsection{Results}\label{sec:summativeresults}
A total of 4932 users were selected as the final set. The drop in users from the starting group of 5025 users can be attributed to one (or more) of the reasons explained in Section~\ref{sec:formativeresults}.
The minimum account age for the sample is 162 days, and the maximum account age is 4267 days. The account age for the users was consistent with the values obtained in the previous experiment. 


The minimum number of followers for any user was 12, and the maximum number for any user was 228033. The average number of followers, including outliers, was 1288 users, while the average number of followers excluding the top 10\% of users was 620. The results obtained are consistent with the formative experiment. 
The number of followers for most of the users in the sample was updated in the 30 days. 4529 users of the entire set of 4932 users had their followers updated. 
The difference in the size of followers (both increase and decrease) was calculated, and users with significant changes to the size of the followers were identified. A significant change is defined as a change in the number of followers of more than 10\% of the follower group (higher than 10) or 100 followers compared to the formative assessment. A total of 313 users were identified with a significant change in their follower population. This measure can identify the potential impact of the change in the initial audience of a tweet towards its performance. 

\begin{figure}[h!]
\centering
\includegraphics[width =\linewidth]{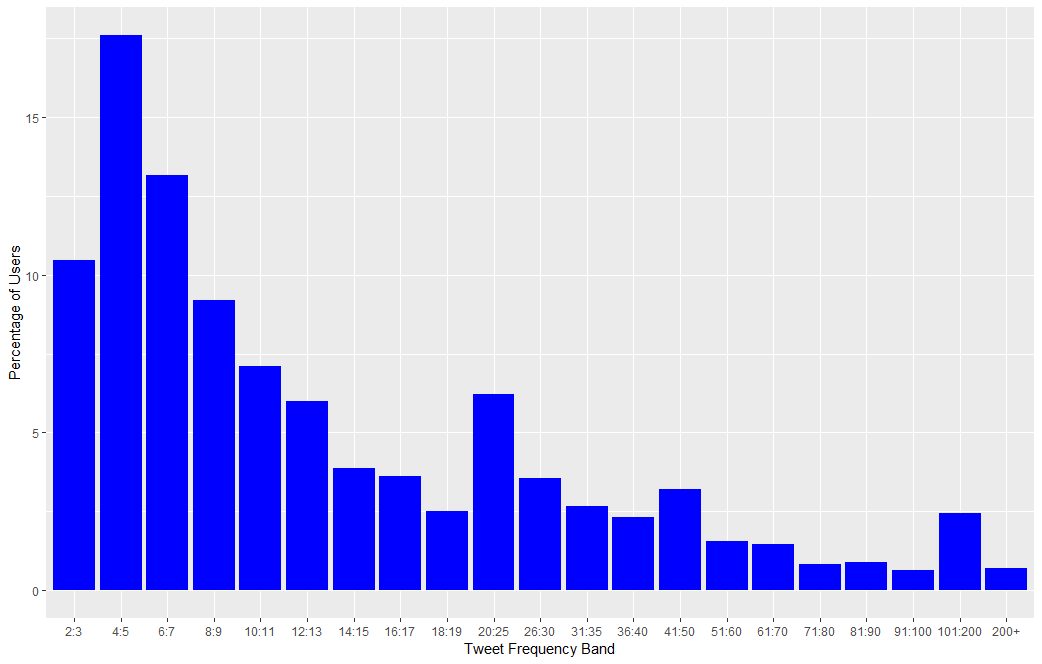}
\caption{Summative -- Original Tweet Frequency Band}
\label{Figure:25}

\end{figure}

Figure~\ref{Figure:25} shows the population distribution of the original Tweet Frequency band of the entire user sample.
Comparing the dataset for the formative assessment in Figure~\ref{Figure:10}, it can be seen that the population in the first six bands increased compared to the other bands. This increase in tweet frequency was expected as the time window for the second sample is relatively smaller (4 Weeks) compared to the formative experiment. A total of 3543 users had their respective tweet frequency bands updated for this experiment. 

\begin{figure*}[ht]
        \centering
        \begin{subfigure}[b]{0.45\textwidth}
            \centering
            \includegraphics[width=\textwidth]{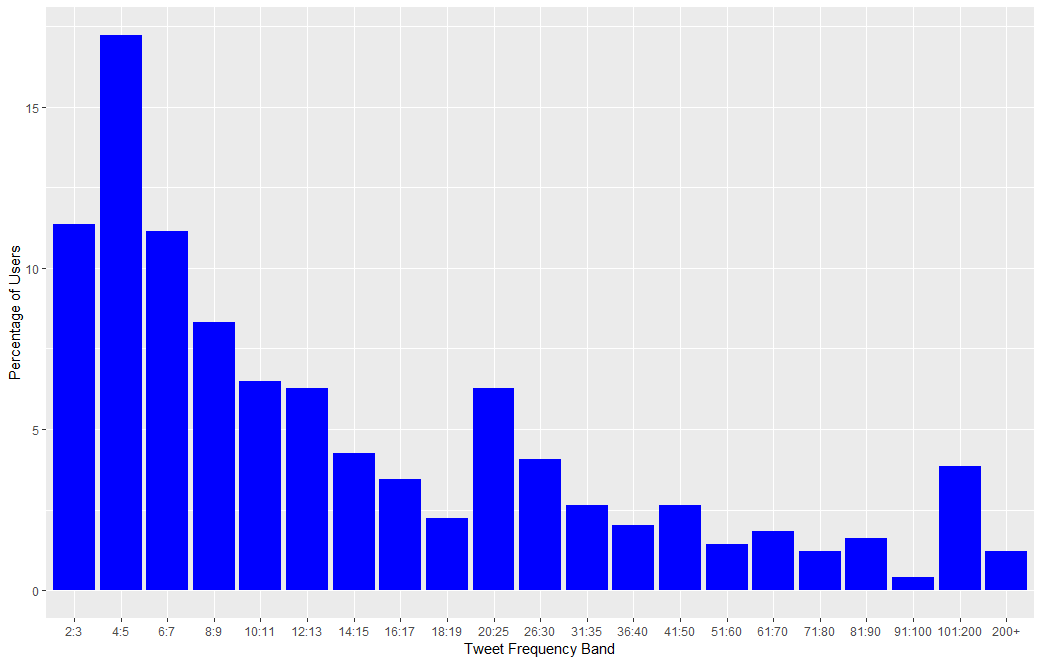}
            \caption[Summative -- AvgTS: Top Performer]%
            {{\small Summative -- AvgTS: Top Performer}}    
            \label{Figure:18}
        \end{subfigure}
        \hfil
        \begin{subfigure}[b]{0.45\textwidth}  
            \centering 
            \includegraphics[width=\textwidth]{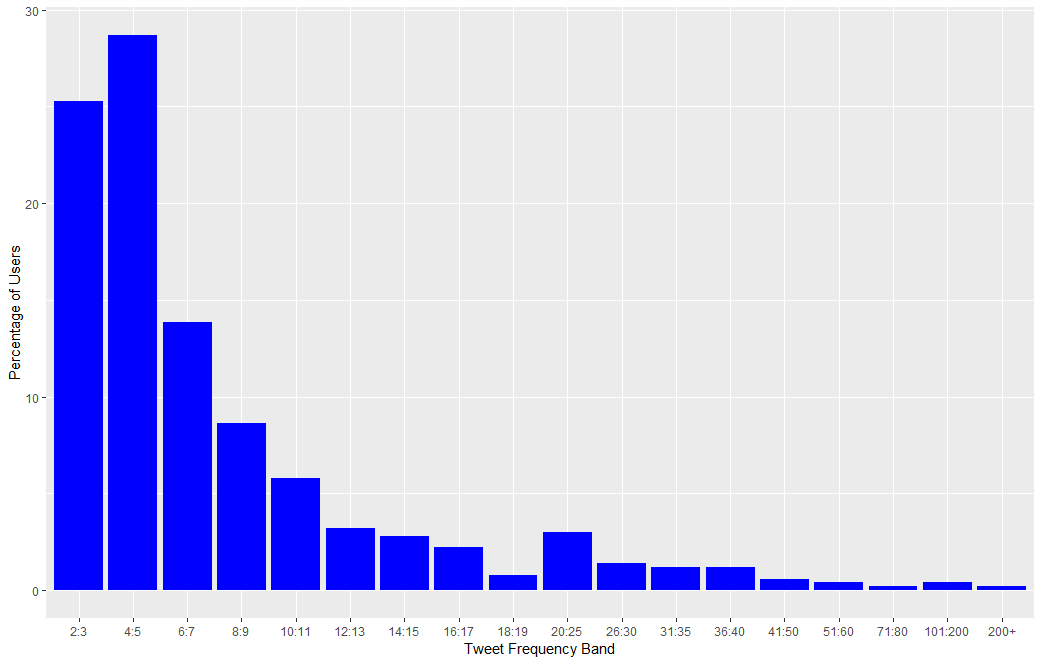}
            \caption[Summative -- AvgAudInpW: Top Performer]%
            {{\small Summative -- AvgAudInpW: Top Performer}}    
            \label{Figure:20}
        \end{subfigure}
        \vskip\baselineskip
        \begin{subfigure}[b]{0.45\textwidth}   
            \centering 
            \includegraphics[width=\textwidth]{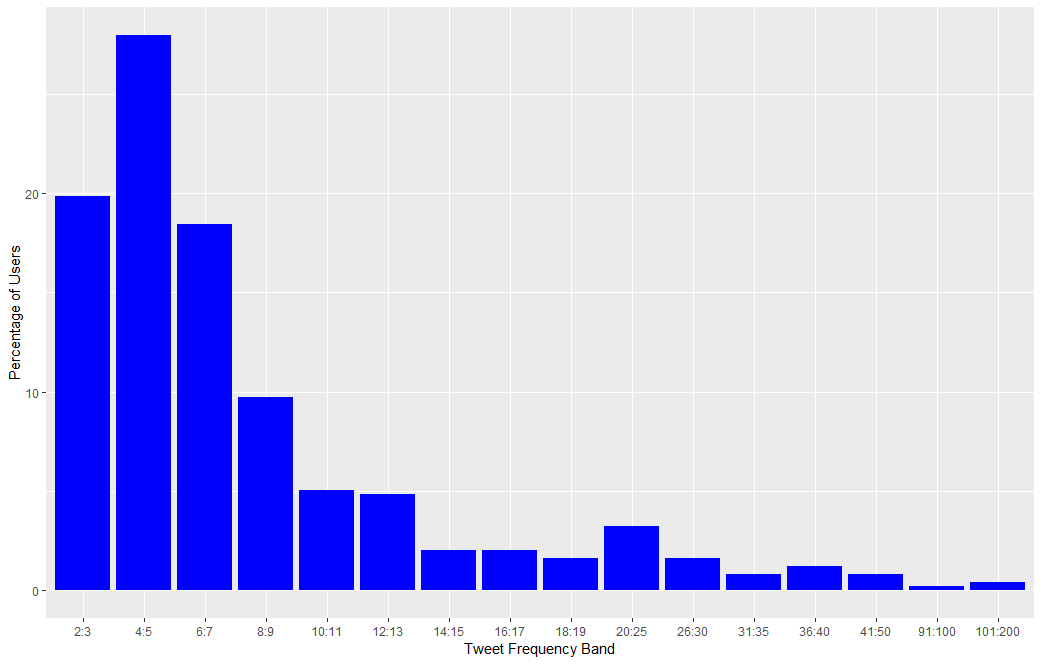}
            \caption[Summative -- AvgTSPc: Top Performer]%
            {{\small Summative -- AvgTSPc: Top Performer}}    
            \label{Figure:21}

        \end{subfigure}
        \hfil
        \begin{subfigure}[b]{0.45\textwidth}   
            \centering 
            \includegraphics[width=\textwidth]{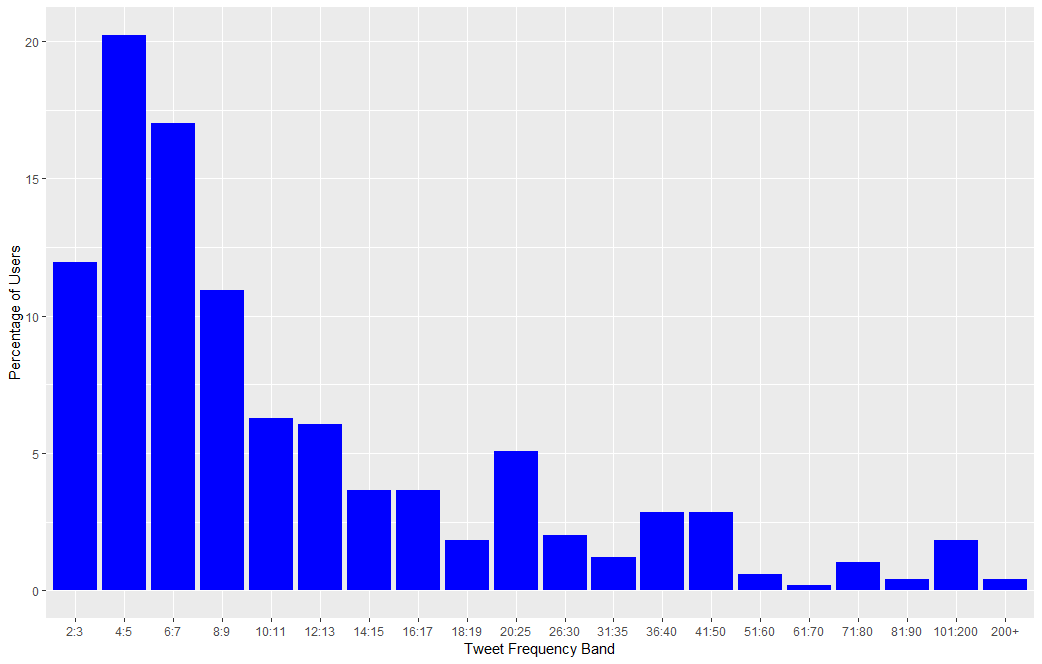}
            \caption[Summative -- \% of Scored Tweets: Top Performer]%
            {{\small Summative -- \% of Scored Tweets: Top Performer}}    
            \label{Figure:19}

        \end{subfigure}
        \caption[ Summative Top Performer ]
        {\small Summative Top Performer} 
        \label{fig:mean and std of nets3}
    \end{figure*}

As we did in the formative experiments, the tweet-frequency band of the users were compared for the top performing users for each metric to determine the relationship between Tweet-frequency and the importance of Tweet. 
The third and 90th percentile AvgTS scores for the sample were 8.044 and 28.133, respectively. Figure 
\ref{Figure:18} represents the Tweet Frequency distribution of the 
top 10\% population. Comparing this plot with the base Tweet Frequency plot in the formative experiments, the distribution is very similar to the base tweet frequency plot, with a slightly higher population in the 20:25 and 101:200 bands. The average number of followers for the two groups was 503 and 546 users, respectively. The Two frequency distribution was consistent with the results obtained in the Formative assessment. 
Based on the distribution for the two groups, it can be inferred that lower tweet frequency groups perform significantly better than the high-frequency groups. It can be attributed to the fact that high-frequency tweeters often have many tweets with a low score which drives the average score down. 

The results' significance was evaluated by performing T-Tests on the summative datasets. Two T-tests were selected for the evaluation: One-Sample T-Test and Welch Two Sample T-Test. The One-Sample T-Test was used to evaluate the top quartile group of the AvgTS against the entire summative sample. The null hypothesis for the test stated that the mean of the Tweet Frequency of the top quartile sample is greater than the mean of the Tweet Frequency of the summative sample. The test results were  
\textit{t = -1.8695, df = 1233, p-value = 0.03089} and since the p-value obtained was less than $\mu_0$ (0.05), the null hypothesis was rejected, and the alternative hypothesis was accepted. Thus it can be concluded that the Tweet Frequency of the top quartile is significantly less than the entire population. 
The Welch Two Sample T-Test was performed on the top quartile for AvgTS for both the Formative and Summative Groups. The null hypothesis for the test stated that the difference in mean of the Tweet Frequency of the two top quartile groups for AvgTS is greater than zero. Since the results are \textit{t = -3.3967, df = 2490.7, p-value = 0.0003463} and the p-value obtained is lesser than 0.05 thus the null hypothesis was rejected, and the alternative hypothesis was accepted. The hypothesis helps to establish that the results obtained during both experiments are consistent with each other while concluding that the top quartile group has a lower mean tweet frequency, i.e. users with a higher AvgTS Score are more likely to be of a lower Tweet Frequency band.

The 90th percentile score for \textbf{prST }is 82.48\%. Figure~\ref{Figure:19} represents the Frequency Band population distribution for prST for the top performer group in the summative sample. The result can be interpreted as 82.48\%, or more of the tweets published by the users in the top performer group were able to engage at least a single user through likes or retweets. The lower Tweet Frequency bands comprise a significant proportion of the population. Similar to the AvgTS metric, the first four frequency bands have the highest percentage of the population. The prST Frequency distribution is consistent with the distribution achieved in the formative experiment. 

Similar to AvgTS, One Sample T-Test and Welch Two Sample T-Test was performed for the top performing group for prST. T-test results show that (\textit{t = -5.0131, df = 493, p-value = 3.742e-07}) the top performer group for prST had a significantly lesser tweet frequency than the entire summative group. The The Welch Two Sample test shows that \textit{t = -5.9038, df = 961.29, p-value = 2.461e-09} the results obtained in the formative and summative experiments were consistent.





The third quartile and 90th percentile score for \textbf{AvgAudInpW} are 0.007 and 0.013 respectively. Figure~\ref{Figure:20} represents the Frequency Band population distribution for AvgAudInpW in the top performer group.
Comparing the plots with the base frequency plot, it can be observed that there is a significant increase in the proportion of users in the first four bands compared to the higher frequency bands. The top performer group has a negligible percentage of uses tweeting greater than 40 tweets per week, with some Tweet frequency bands missing from the group. The average size of the follower group for the top performer group was 321, which is significantly lower than the sample average, indicating a negative relationship between the metric and the follower group size, as expected. The results are consistent with the results of the Formative Experiment for the same metric. 
One Sample T-Test on the top quartile group for the AvgAudInpW against the mean of the summative sample
shows that (\textit{t = -19.67, df = 1279, p-value < 2.2e-16}) 
the tweet frequency of the top quartile is significantly less than the summative sample. Furthermore, the Welch Two Sample T-Test on the top quartile of the Formative and Summative groups for AvgAudInpW shows that 
(\textit{t = -2.6037, df = 2456.8, p-value = 0.004639}) 
the results for AvgAudInpW were consistent with those obtained for the previous metrics.

The third quartile and 90th percentile score for \textbf{AvgTSPc} are 37.85\% and 50.64\% respectively. The scores are slightly higher than the Formative study, as expected, due to the smaller size of the summative dataset. Figure~\ref{Figure:21} represents the Frequency Band population distribution for AvgTSPc for the summative sample in the top performer group.


The average number of followers for the top performer group was 452, whereas the average number of tweets evaluated per users in the group was 207. Both these values were consistent with the complete summative sample indicating the metric is not dependent on these attributes.  The results obtained for the frequency distribution are consistent with the formative assessment. 
One Sample and Welch Two Sample T-tests were evaluated for the top performer group of the AvgTSPc metric. One sample test shows that (\textit{t = -16.732, df = 1232, p-value < 2.2e-16}) the tweet frequency of the top quartile was significantly less than the summative sample. The Welch test shows that (\textit{t = -6.8046, df = 2395.8, p-value = 6.377e-12}) top quartile group had a lower tweet frequency. 




We also investigated the effect of a significant change in followers on performance. The user-level metrics Average Tweet score and prST were compared for the 313 users with a significant change in followers to establish a possible relationship between the metrics and the change in followers. However, there was no discernible pattern identifiable through the metric.

\section{Celebrity Accounts}
The formative and summative experiments detailed above consider the non-verified account users in their sample. 
The experiments were exclusively carried out for top-rated celebrity accounts to study the variation of the defined metrics for verified accounts, and to establish the relationship of Tweet frequency with the popularity of the tweets for this account. 
The \textbf{participants} for the experiment were selected based on their popularity, i.e., the number of followers. The leading ten political and 10 celebrity accounts in the UK were selected as shown in Table \ref{table:9}. 
The historical tweet dataset and user profile attributes for each of the 20 accounts were collected. 
The \textbf{tools and materials} used and the experimental \textbf{design} for this experiment were unchanged from the previously described experiments. 
\subsection{Results}
Table~\ref{table:9} describes the user profile demographics for the 20 selected accounts. The Table enlists the user's followers, account age, and average original tweet frequency per week (AvgOrTpW). The users for this experiment were not divided into frequency bands as the audience size was small. The user-level metrics were evaluated against the absolute frequency of tweeting for each user.

\begin{table*}[]
\footnotesize
\begin{tabular}{|l|l|l|l||l|l|l|l|}
\hline
\textbf{user\_name }    & \textbf{followers} & \textbf{age}  & \textbf{avgOrTweetWeek} & \textbf{user\_name}       & \textbf{followers} & \textbf{age}  & \textbf{avgOrTweetWeek} \\ \hline
David\_Cameron* & 1884293   & 3140 & 10             & Coldplay         & 23953837  & 3502 & 15             \\ \hline
jeremycorbyn*   & 1844646   & 3091 & 33             & Nigel\_Farage*    & 1209391   & 3499 & 22             \\ \hline
UKLabour*      & 615414    & 3786 & 33             & SadiqKhan*        & 943412    & 3490 & 39             \\ \hline
MayorofLondon*  & 3226290   & 3750 & 42             & onedirection     & 30906220  & 2846 & 14             \\ \hline
LiamPayne      & 32305743  & 2976 & 12             & WayneRooney      & 16857963  & 2673 & 6              \\ \hline
NicolaSturgeon* & 915304    & 2968 & 23             & Ed\_Miliband*     & 745943    & 3301 & 12             \\ \hline
EmmaWatson     & 28543868  & 2953 & 4              & ShamaJunejo*      & 197825    & 3297 & 111            \\ \hline
zaynmalik      & 27202998  & 2927 & 9              & theresa\_may*     & 598012    & 777  & 7              \\ \hline
Harry\_Styles  & 32288027  & 2914 & 11             & Louis\_Tomlinson & 32900121  & 3219 & 10             \\ \hline
Adele          & 27924275  & 2906 & 2              & edsheeran        & 19508485  & 3214 & 19             \\ \hline
\end{tabular}
\caption{Celebrity Account Demographics. * shows the selected political accounts}
\label{table:9}
\end{table*}

\begin{figure*}[ht]
        \centering
        \begin{subfigure}[b]{0.45\textwidth}
            \centering
            \includegraphics[width=\textwidth]{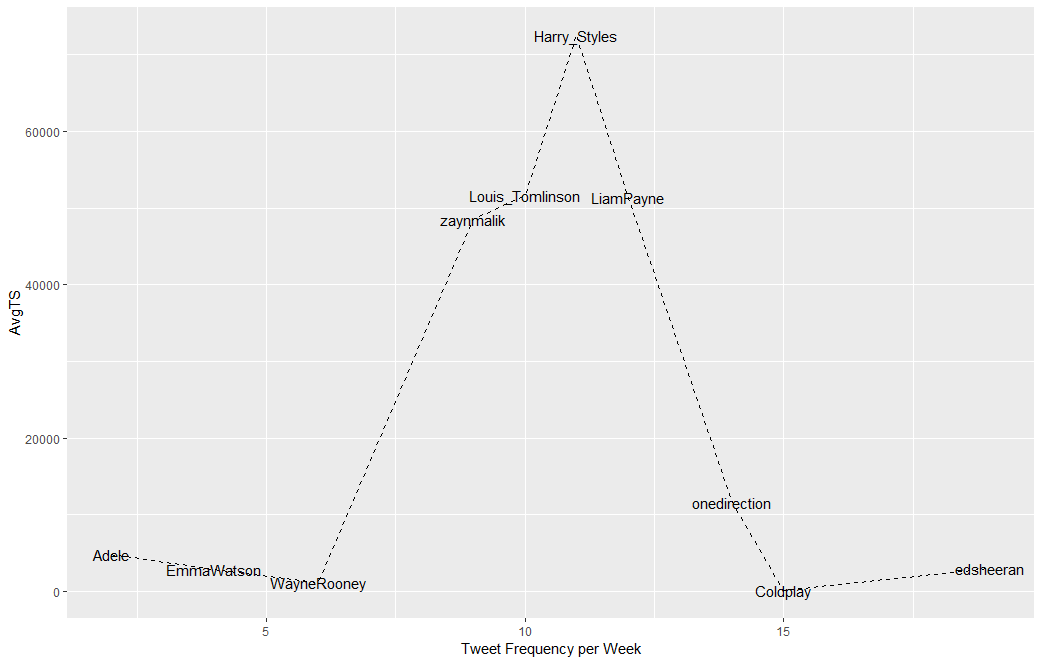}
            \caption[AvgTS: Celebrity account]%
            {{\small AvgTS: Celebrity account}}    
            \label{Figure:22a}
        \end{subfigure}
        \hfil
        \begin{subfigure}[b]{0.45\textwidth}  
            \centering 
            \includegraphics[width=\textwidth]{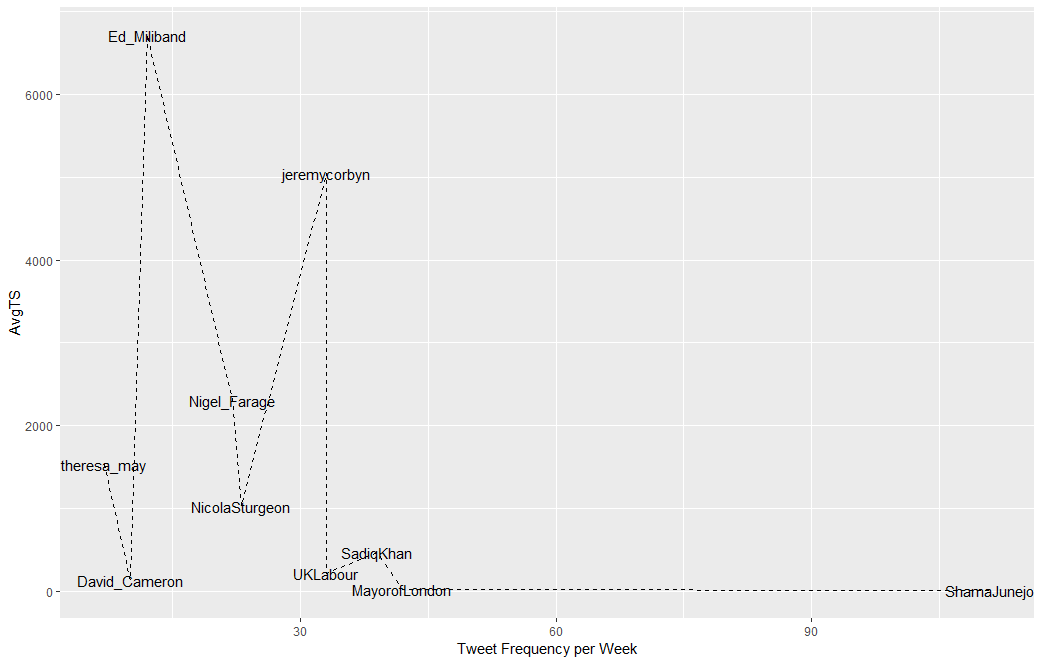}
            \caption[AvgTS: Political account]%
            {{\small AvgTS: Political account}}    
            \label{Figure:22b}
        \end{subfigure}
        \vskip\baselineskip
        \begin{subfigure}[b]{0.45\textwidth}   
            \centering 
            \includegraphics[width=\textwidth]{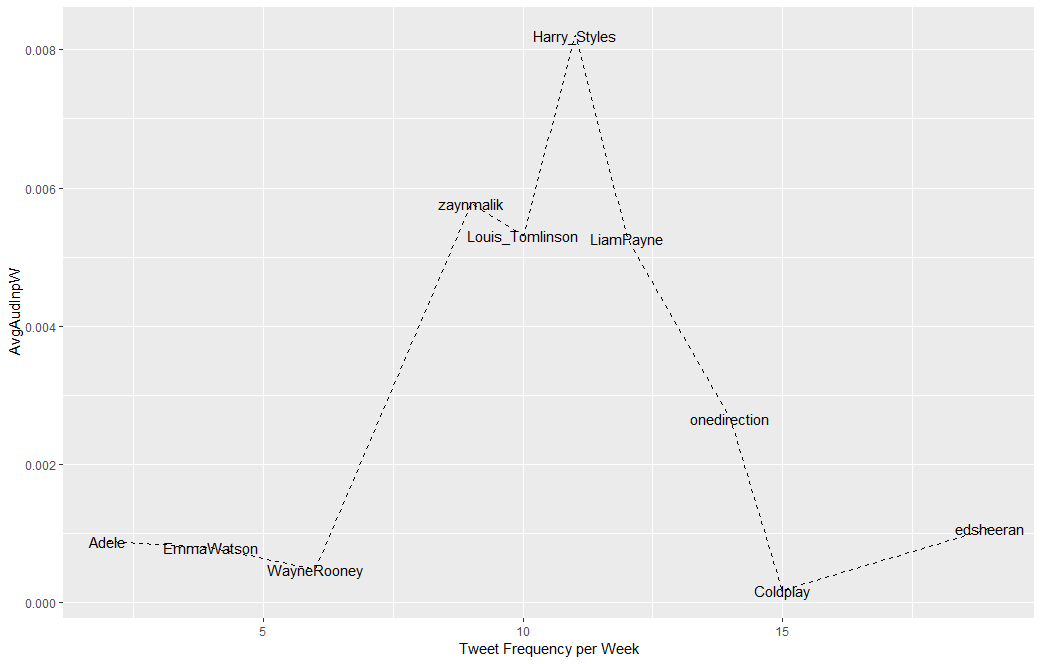}
            \caption[AvgAudInpW : Celebrity account]%
            {{\small AvgAudInpW : Celebrity account}}    
            \label{Figure:23a}

        \end{subfigure}
        \hfil
        \begin{subfigure}[b]{0.45\textwidth}   
            \centering 
            \includegraphics[width=\textwidth]{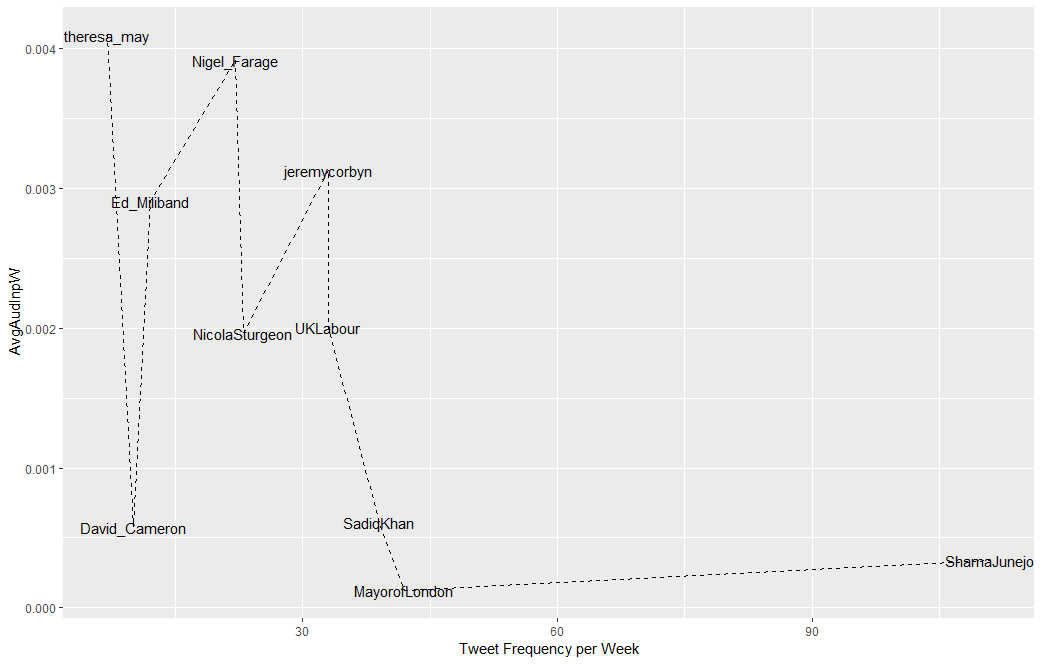}
            \caption[AvgAudInpW: Political account]%
            {{\small AvgAudInpW: Political account}}    
            \label{Figure:23b}

        \end{subfigure}
        \caption[ Celebrity Accounts: AvgTS and AvgAudInpW ]
        {\small Celebrity Accounts: AvgTS and AvgAudInpW} 
        \label{fig:mean and std of nets}
    \end{figure*}

Figure~\ref{Figure:22a} and~\ref{Figure:22b} represent the \textbf{AvgTS} distribution for the Celebrity and Political accounts respectively. Similar to the Formative and summative experiments, it can be observed that the score tends to decrease with an increase in tweet frequency for both sets of users. 
The \textbf{prST} metric does not provide any meaningful insight into the importance of the tweets as almost all users achieve a 100\% score for this measure. As the accounts being evaluated were the top accounts in their respective fields, almost all tweets published by them were able to engage at least a single user through retweets or likes. 
Figure \ref{Figure:23a} and \ref{Figure:23b} represent the distribution of \textbf{AvgAudInpW} for the Celebrity and Political Accounts respectively. The Figure~\ref{Figure:23a} plot for the Celebrity account is consistent with the plot for the AvgTS for the same users. The users with extremely low frequency are less able to drive engagement from their audience, although the engagement tends to decrease after a certain threshold limit. For this set, the users with 7-12 tweets per week drive maximum engagement.
Similarly, for the Political set of users, the lower tweet frequency users can drive higher audience interaction than those with higher frequency. Although compared with the AvgTS plot, it can be observed that the score of some users has increased significantly compared to the others. It can be attributed to the significantly lower number of followers for these users, indicating the negative relationship between the metric and follower group size, as established in the previous section.

\section{Discussion}\label{sec:discussion}
The importance of a tweet or an author is highly subjective where the meaning of importance or the level of importance can vary significantly with change in the context of the reader. This study views importance as a measure towards providing a solution to the problem of information overload on Twitter. The importance of a tweet is thus defined as its ability to engage its reader through some action. 
Based on our experiments it can be concluded that the top performers for all proposed metrics of importance are constituted by a higher proportion of low-frequency tweeters. 
The (pseudo) random user sample considered consisted of a wide range of tweet frequency authors although a high percentage of the sample was part of the lower tweet frequency bands. Thus to establish the relationship, a change in tweet frequency distribution was observed. While selecting the tweets evaluated for the experiment, Tweet characteristics based on tweet text, the existence of image or URL, the existence of user mentions, time of publishing, or other possible characteristics which may contribute to the popularity of the tweet were not considered. All Tweets published by the users regardless of any external/internal attributes were evaluated against the same metrics. This was done to ensure that the results obtained are not affected by any external factors, and the tweets present a realistic sample of the Twitter ecosphere. Although the experiments establish low tweet frequency as a characteristic of users perceived to be important, it should be noted that it does not establish a causal relationship between tweet frequency and the importance of tweets.

The outcomes and insights of our research can have multiple \textbf{applications}. For instance, it can be used for tweet reordering. The Tweet-level and User-level metrics were identified to provide a useful measure to compare the tweets and their authors. The home timeline of users can be modified by reordering the Tweets published by the perceived importance of the authors. 
Finally, the proposed metrics could be used to create a scoring index to rank the users globally to identify top users who are perceived to be important by their followers. Such ranking could help users to identify users whom they could follow. 

Our study is not without \textbf{limitations}. It was assumed that all followers or the entire initial audience of a user read every tweet posted by the author which is almost never right and the results are biased against primarily two sets of users: users who have a large following of dormant or inactive user, and users that are listed low on the home timeline by the current ordering rules of the Twitter systems. For both, the tweets are read only by a portion of their followers thus evaluating against the entire audience provides low scores for such users. Ideally, the tweets should be evaluated based on their read counts rather than the number of followers. As Twitter does not track the reading activity of a tweet, it is not possible to evaluate the tweets based on that metric. A further limitation of using followers as the initial audience rather than read count was that as the number of followers of a user was not stored at each instance, a tweet was published, the current number of followers of a user was considered as the initial population for the entire dataset. 

\section{Conclusion}\label{sec:conclusion}
This research tackles the problem of information overload by measuring the importance of tweets and their authors. The importance of a Tweet in the context of this research is considered the ability of a tweet to engage its reader by initiating action by the reader, and two Tweet level measures of importance are proposed -- Tweet Score and Tweet Score Percentile.
The hypothesis ``low-frequency tweeters have more to say'' was tested by evaluating the top performer groups for the proposed metrics. Through the results obtained, a relationship between the frequency of tweeting and the importance of tweets was established. It was observed that users tweeting less than ten tweets per week were more likely to be important than those tweeting more than ten tweets per week. 
Thus the proposed measures of importance along with the Tweet frequency of the author were established as useful metrics for identifying important users.

\bibliographystyle{cas-model2-names}

\bibliography{refs}




\end{document}